\documentclass[aps,prd,superscriptaddress,twocolumn,showpacs]{revtex4}
\usepackage{graphicx}
\usepackage{epstopdf}
\usepackage{amsmath}
\usepackage{amsfonts}
\usepackage{amssymb}
\usepackage{latexsym}

\begin{document}
\title{Remarks on a $B \wedge F$ model with topological mass from gauging spin}
\author{Patricio Gaete} \email{patricio.gaete@usm.cl} 
\affiliation{Departamento de F\'{i}sica and Centro Cient\'{i}fico-Tecnol\'ogico de Valpara\'{i}so-CCTVal, Universidad T\'{e}cnica Federico Santa Mar\'{i}a, Valpara\'{i}so, Chile}
\author{Jos\'{e} A. Helay\"{e}l-Neto}\email{helayel@cbpf.br}
\affiliation{Centro Brasileiro de Pesquisas F\'{i}sicas (CBPF), Rio de Janeiro, RJ, Brasil} 
\date{\today}

\begin{abstract}
Aspects of screening and confinement are reassessed for a $B \wedge F$ model with topological mass with the gauging of spin. Our discussion is carried out using the gauge-invariant, but path-dependent, variables formalism. We explicitly show that the static potential profile is the sum of a Yukawa and a linear potential, leading to the confinement of static external charges. Interestingly enough, similar results are obtained in a theory of antisymmetric tensor fields that results from the condensation of topological defects as a consequence of the Julia-Toulouse mechanism.
\end{abstract}
\maketitle

\section{Introduction}

The mass generation mechanism in gauge theories is one of the most striking and fascinating quantum phenomenon, which has been of great interest since the work of St\"uckelberg \cite{Stuckelberg}. Mention should be made, at this point, to the renowned Higgs mass generation mechanism, which was theoretically predicted in 1964 \cite{Englert,Higgs1,Higgs2} and experimentally confirmed in 2012 \cite{Atlas,CMS}.

In this connection, it becomes of interest, in particular, to recall the Schwinger model or Quantum Electrodynamics in $(1+1)$ dimensions \cite{Schwinger} due to some very interesting and peculiar properties that it possesses, like fermion confinement and the energy spectrum contains a massive mode in spite of the gauge invariance of the original Lagrangian. We also recall that $B \wedge F$ models, a Chern-Simons-like formulation with antisymmetric tensor fields, also experience mass generation \cite{EuroSma, Harikumar}.

Meanwhile, in previous works \cite{GaeteWot1,GaeteWot2}, we have considered the confinement versus screening properties of some theories of massless antisymmetric tensors, magnetically and electrically coupled to topological defects that eventually condensate, as a consequence of the Julia-Toulouse mechanism \cite{Toulouse,Quevedo}. In passing, we recall that this mechanism is the dual to the Higgs mechanism and has been shown to lead a massive antisymmetric theories with a jump rank. To be more precise, we have illustrated that in the presence of two tensor fields the condensation induces not only a mass term and a jump of rank but also a $B \wedge F$ coupling which is responsible for the change from screening to the confining phase of the theory. More recently, we have computed the interaction energy between two test charges for a $(3+1)$-dimensional generalization of the bosonized Schwinger model \cite{Auriliaetal}. Our results show that the static potential profile contains a linear term leading to the confinement of probe charges, exactly as in the original model in $(1+1)$ dimensions. It should be further noticed that the same $4$-dimensional model also appears as one version of the $B \wedge F$  models in $(3+1)$ dimensions under dualization of St\"uckelberg-like massive gauge theories \cite{EuroSma}. Interestingly, this particular model is characterized by the mixing between a $U(1)$ field and an Abelian $3$-form field of the type that appears in the topological sector of quantum chromodynamics (QCD).

On the other hand, we further note that recently a novel way to induce the $B \wedge F$ term has been considered in \cite{Diamantini}. The crucial ingredient of this development is that the $B \wedge F$-term is generated at one loop by coupling the antisymmetric gauge field to the vorticity of charged fermions.

With these considerations in mind, the purpose of this work is to further elaborate on the physical content of
this $(3+1)$-dimensional $B \wedge F$ effective action. To do this, we shall work out the static potential for the
theory under consideration along the lines of \cite{GaeteWot1,GaeteWot2}. The advantage of using this development lies in the fact that the interaction energy between two static charges is obtained once a judicious identification of the physical degrees of freedom is made \cite{Pato1}. 

 Our work is organized as follows: in Section II, we shall reexamine the calculation of the interaction energy between static point-like sources for conventional $B \wedge F$ models by using the gauge-invariant, but path-dependent variables formalism.
In Section III, we extend our analysis for a $B \wedge F$ model with topological mass from gauging spin.
Interestingly enough, for this effective model, the static potential profile contains a linear term, leading to the confinement of static charges. Finally, some concluding remarks are made in Sec. IV. 

In our conventions the signature of the metric is ($+1,-1,-1,-1$).

\section{Brief review on $B \wedge F$ models} 

We start our analysis by reconsidering the calculation of the interaction energy between static point-like sources for $ B \wedge F$ models by using the gauge-invariant but path-dependent variables formalism along the lines of \cite{GaeteWot1, GaeteWot2}. To do this, we shall compute the expectation value of the energy operator $H$ in the physical state $\left| \Phi  \right\rangle $ describing the sources, which we shall denote by ${\left\langle H \right\rangle _\Phi }$.

Before we proceed to work out explicitly the energy, we shall describe very briefly $ B \wedge F$ models. In such a case we consider the Lagrangian density:
\begin{equation}
{\cal L} =  - \frac{1}{4}F_{\mu \nu }^2\left( A \right) + \frac{1}{{12}}H_{\mu \nu \rho }^2\left( B \right) 
+ \frac{m}{{24}}{\varepsilon ^{\mu \nu \rho \sigma }}{B_{\mu \nu }}{\partial _{[\rho }}{A_{\sigma ]}}. \label{BF05}
\end{equation}
At this point, it is advantageous to recall the relevant aspects of the analysis described previously \cite{Auriliaetal}. The first observation is that the Lagrangian density (\ref{BF05}) can be brought into the form 
\begin{equation}
{{\cal L}} =  - \frac{1}{4}F_{\mu \nu }^2 - \frac{1}{2}{\tilde H_\sigma }{\tilde H^\sigma } 
- \frac{m}{6}{\tilde H^\sigma }{A_\sigma }, \label{BF10}
\end{equation}
where  ${\tilde H^\mu } = 
{\raise0.5ex\hbox{$\scriptstyle 1$}\kern-0.1em/\kern-0.15em\lower0.25ex\hbox{$\scriptstyle 2$}}{\varepsilon ^{\mu \nu \lambda \rho }}{\partial _\nu }{B_{\lambda \rho }}$.
Next, in order to eliminate the dual-field $\tilde H^{\sigma}$ in favor of the $A_{\mu}$-field, we should not forget that $\tilde H^{\mu}$ satisfies the constraint ${\partial _\mu }{\tilde H^\mu } = 0$. So, we introduce a Lagrange multiplier, $\chi$, to take the constraint into account (this constraint is actually the Bianchi identity associate to ${H_{\mu \nu \kappa }}$). By considering this last constraint, the Lagrangian density (\ref{BF10}) becomes  
\begin{equation}
{{\cal L}} = - \frac{1}{4}F_{\mu \nu }^2 - \frac{1}{2}{\tilde H_\sigma }{\tilde H^\sigma }
- \frac{m}{6}{\tilde H^\sigma }{A_\sigma } + \chi {\partial _\sigma }{\tilde H^\sigma }. \label{BF15}
\end{equation}
In such a situation, and letting ${Z_\sigma } \equiv {A_\sigma } + \frac{6}{m}{\partial _\sigma }\chi$, with $  {Z_{\mu \nu }} = {F_{\mu \nu }}$, equation  (\ref{BF15}) reduces to
\begin{equation}
{{\cal L}} =  - \frac{1}{4}Z_{\mu \nu }^2 - \frac{1}{2}{\tilde H_\sigma }{\tilde H^\sigma } 
- \frac{m}{6}{\tilde H^\sigma }{Z_\sigma }.\label{BF20} 
\end{equation}
Then equation (\ref{BF05}) can be written in the form
\begin{equation}
{{\cal L}} =  - \frac{1}{4}Z_{\mu \nu }^2 + \frac{1}{2}{\mu ^2}Z_\mu ^2, \label{BF25}
\end{equation}
where we have made use of ${W_\sigma } \equiv {\tilde H_\sigma } + \frac{m}{6}{Z_\sigma }$ and
 ${\mu ^2} \equiv {\raise0.5ex\hbox{$\scriptstyle {{m^2}}$}\kern-0.1em/\kern-0.15em
\lower0.25ex\hbox{$\scriptstyle {36}$}}$. Thus, finally we end up with a Maxwel-Proca theory.

Accordingly, this effective model provides us with a suitable starting point to study the interaction energy. Next, we also notice that before proceeding with the determination of the energy, it is necessary to restore the gauge invariance in (\ref{BF25}). Following an earlier procedure \cite{GaeteWot1,GaeteWot2}, we may express equation (\ref{BF25}) as
\begin{equation}
{{\cal L}_{eff}} =  - \frac{1}{4}{F_{\mu \nu }}\left( {1 + \frac{{{\mu ^2}}}{\Delta }} \right){F^{\mu \nu }}, \label{BF30}
\end{equation}
where $\Delta  = {\partial _\mu }{\partial ^\mu }$.

Having established the new effective Lagrangian, we can now compute the interaction energy. To this end, we first
consider the Hamiltonian framework of this new effective theory. The canonical momenta are found to be ${\Pi ^\mu } =  - \left( {1 + \frac{{{\mu ^2}}}{\Delta }} \right){F^{0\mu }}$. This yields the usual primary constraint ${\Pi ^0} = 0$, and 
${\Pi ^i} =  - \left( {1 + \frac{{{\mu ^2}}}{\Delta }} \right){F^{0i}}$. Therefore the canonical Hamiltonian is
\begin{eqnarray}
{H_C} &=&\int {{d^3}x} \left\{ { - {A_0}{\partial _i}{\Pi ^i} - \frac{1}{2}{\Pi _i}{{\left( {1 + \frac{{{\mu ^2}}}{\Delta }} \right)}^{ - 1}}{\Pi ^i}} \right\} \nonumber\\
&+& \int {{d^3}x} \left\{ {  \frac{1}{4}{F_{ij}}{F^{ij}}} \right\}. \label{BF35} 
\end{eqnarray}
Temporal conservation of the primary constraint ${\Pi _0}$ leads to the secondary constraint ${\Gamma _1}\left( x \right) \equiv {\partial _i}{\Pi ^i} = 0$. It can be easily seen that there are no further constraints in the theory. According to the general theory, we obtain the extended Hamiltonian by adding all the first-class constraints with arbitrary constraints. We thus write $H = {H_C} + \int {{d^3}x} \left( {{c_0}\left( x \right){\Pi _0}\left( x \right) + {c_1}\left( x \right){\Gamma _1}\left( x \right)} \right)$, where ${c_0}(x)$ and ${c_1}(x)$ are the Lagrange multipliers. Moreover, it follows from this Hamiltonian that ${\dot A_0}(x) = \left[ {{A_0}(x),H} \right] = {c_0}(x)$, which is an arbitrary function. Since $\Pi_0=0$, neither $A^0$ nor $\Pi^0$ are of interest in describing the system and may be discarded from the theory. In fact, the term containing $A_0$ is redundant, because it can be absorbed by redefining the function ${c^ \prime }\left( x \right)$. Therefore, the Hamiltonian is now given as
\begin{eqnarray}
{H} &=&\int {{d^3}x} \left\{ {  - \frac{1}{2}{\Pi _i}{{\left( {1 + \frac{{{\mu ^2}}}{\Delta }} \right)}^{ - 1}}{\Pi ^i}} 
+\frac{1}{4}{F_{ij}}{F^{ij}}\right\} \nonumber\\
&+& \int {{d^3}x} \left\{ c^ \prime ({ {\partial _i}{\Pi ^i} })\right\}  , \label{BF40} 
\end{eqnarray}
where ${c^ \prime }\left( x \right) = {c_1}\left( x \right) - {A_0}(x)$.

Since there is one first-class constraint ${\Gamma _1}\left( x \right)$, we choose one gauge condition that makes the full set of constraints to become second-class. A particularly convenient choice is
\begin{equation}
\Gamma _2 \left( x \right) \equiv \int\limits_{C_{\xi x} } {dz^\nu
} A_\nu \left( z \right) \equiv \int\limits_0^1 {d\lambda x^i }
A_i \left( {\lambda x} \right) = 0. \label{BF45}
\end{equation}
where  $\lambda$ $(0\leq \lambda\leq1)$ is the parameter describing
the spacelike straight path $ z^i = \xi ^i  + \lambda \left( {x -\xi } \right)^i $, and $ \xi $ is a fixed point (reference point). There is no essential loss of generality if we restrict our considerations to $ \xi ^i=0 $. As a consequence, the only nontrivial Dirac bracket for the canonical variables is given by
\begin{eqnarray}
\left\{ {A_i \left( x \right),\Pi ^j \left( y \right)} \right\}^ *
&=& \delta _i^j \delta ^{\left( 3 \right)} \left( {x - y} \right) \nonumber\\
&-&\partial _i^x \int\limits_0^1 {d\lambda x^i } \delta ^{\left( 3
\right)} \left( {\lambda x - y} \right). \label{BF50}
\end{eqnarray}

We are now in the position to calculate the interaction energy for the model under consideration. For this purpose, we shall compute the expectation value of the energy operator $H$ in the physical state $\left| \Phi  \right\rangle$. In this context, we recall that the physical state $\left| \Phi  \right\rangle$ can be written as \cite{Dirac}: 
\begin{equation}
\left| \Phi  \right\rangle  \equiv \left| {\bar \Psi ({\bf y})\Psi ({{\bf y}^ \prime })} \right\rangle  = \bar \psi ({\bf y})\exp (iq\int_{{{\bf y}^ \prime }}^{\bf y} {d{z^i}{A_i}(z)} )\psi ({{\bf y}^ \prime })\left| 0 \right\rangle,                              
\label{BF55}
\end{equation}
where $\left| 0 \right\rangle$ is the physical vacuum state and the
line integral appearing in the above expression is along a spacelike
path starting at ${\bf y}\prime$ and ending at $\bf y$, on a fixed
time slice. 

Taking the above Hamiltonian structure into account, we see that
\begin{eqnarray}
\Pi _i \left( x \right)\left| {\overline \Psi \left( \mathbf{y }\right)\Psi
\left( {\mathbf{y}^ \prime } \right)} \right\rangle &=& \overline \Psi \left( 
\mathbf{y }\right)\Psi \left( {\mathbf{y}^ \prime } \right)\Pi _i \left( x
\right)\left| 0 \right\rangle \nonumber\\
&+& q\int_ {\mathbf{y}}^{\mathbf{y}^ \prime } {\
dz_i \delta ^{\left( 3 \right)} \left( \mathbf{z - x} \right)} \left| \Phi
\right\rangle. \nonumber\\
 \label{BF60}
\end{eqnarray}

Having pointed out this observation, and since the fermions are taken to be infinitely massive (static sources), we can substitute  $\Delta$ by $- {\nabla ^2}$ in equation (\ref{BF40}). Therefore, the interaction energy takes the form
\begin{equation}
{\left\langle H \right\rangle _\Phi } = {\left\langle H \right\rangle _0} + {V_1} , \label{BF65}
\end{equation}
where ${\left\langle H \right\rangle _0} = \left\langle 0 \right|H\left| 0 \right\rangle$. The $V_1$ term is given by
\begin{equation}
{V_1} = \left\langle \Phi  \right|\int {{d^3}x} \left[ { - \frac{1}{2}{\Pi _i}{{\left( {1 - \frac{{{\mu ^2}}}{{{\nabla ^2}}}} \right)}^{ - 1}}{\Pi ^i}} \right]\left| \Phi  \right\rangle.  \label{BF70}
\end{equation}

Following our earlier procedure \cite{GaeteWot1,GaeteWot2}, we see that the static potential takes the form
\begin{equation}
V =  - \frac{{{q^2}}}{{4\pi }}\frac{{{e^{ - \mu L}}}}{L}, \label{BF80}
\end{equation}
where $L = |{\bf y} - {{\bf y}^ {\prime}}|$.

An alternative way of stating the previous result is by considering the expression \cite{Pato1} 
\begin{equation}
V \equiv q\left( {{\cal A}_0 \left( {\bf 0} \right) - {\cal A}_0 \left( {\bf L} \right)} \right), \label{BF85}
\end{equation}
where the physical scalar potential is given by
\begin{equation}
{\cal A}_0 (t,{\bf r}) = \int_0^1 {d\lambda } r^i E_i (t,\lambda
{\bf r}). \label{BF90}
\end{equation}
This equation follows from the vector gauge-invariant field expression
\begin{equation}
{\cal A}_\mu  (x) \equiv A_\mu  \left( x \right) + \partial _\mu  \left( { - \int_\xi ^x {dz^\mu  A_\mu  \left( z \right)} } \right), \label{BF95}
\end{equation}
where the line integral is along a spacelike path from the point $\xi$ to $x$, on a fixed slice time. It should again be stressed here that the gauge-invariant variables (\ref{BF90}) commute with the sole first constraint (Gauss' law), showing in this way that these fields are physical variables. 

We also recall that Gauss' law for the present model reads
\begin{equation}
{\partial _i}{\Pi ^i} = {J^0},  \label{BF100}
\end{equation}
where, $J^0$, is the external source. It should be further recalled that, 
${\bf E} = \frac{{{\nabla ^2}}}{{{\nabla ^2} - {\mu ^2}}}{\bf {\Pi}}$, and for, ${J^0}({\bf x}) = q{\delta ^{\left( 3 \right)}}\left( {\bf x} \right)$, we can express (\ref{BF90}) as
\begin{equation}
{{\cal A}_0}(t,{\bf x}) = \int_0^1 {d\lambda } {x^i}{\left( {1 - \frac{{{\mu ^2}}}{{{\nabla ^2}}}} \right)^{ - 1}}{\Pi _i}\left( {\lambda {\bf x}} \right). \label{BF105}
\end{equation}

With the aid of equations (\ref{BF85}) and  (\ref{BF105}) , we readily find that the interaction energy for a pair of static point-like opposite charges located at $\bf 0$ and $\bf L$ is given by
\begin{equation}
V =  - \frac{{{q^2}}}{{4\pi }}\frac{{{e^{ - \mu L}}}}{L}, \label{BF110}
\end{equation}
after subtracting a self-energy term.

\section{$B \wedge F$ model with topological mass from gauging spin} 

As already stated, our next undertaking is to use the ideas of the previous Section in order to consider $B \wedge F$ model with topological mass with the gauging of spin. For this purpose, the authors of Ref. \cite{Diamantini} consider the four-dimensional space-time Lagrangian density:
\begin{eqnarray}
{\cal L} &=& \bar \psi {\gamma ^\mu }\left( {i{\partial _\mu } + e{A_\mu }} \right)\psi  - m\bar \psi \psi  + g{B_{\mu \nu }}{J^{\mu \nu }} - \frac{1}{4}{F_{\mu \nu }}{F^{\mu \nu }} \nonumber\\
&+& \frac{1}{{12}}{H_{\mu \nu \lambda }}{H^{\mu \nu \lambda }}.
 \label{BF115}
\end{eqnarray}
The crucial idea underlying this suggestion consists in proposing the interaction Lagrangian in the form:
\begin{equation}
{{\cal L}^{Int}}\equiv g{B_{\mu \nu }}{J^{\mu \nu }} = \frac{{2mg}}{\Delta }{F_\mu }{J^\mu }, \label{BF120}
 \end{equation}
where ${F_\mu } = \frac{1}{2}{\varepsilon _{\mu \nu \alpha \beta }}{\partial ^\nu }{B^{\alpha \beta }}$. Given its relevance, it is of interest to study the effect of the above scenario on a physical observable. We also note here that integrating out the fermionic field in (\ref{BF110}) induces an effective model for the $A_{\mu}$- and $B_{\mu\nu}$ - fields. Hence, one gets the following effective Lagrangian density
\begin{eqnarray}
{\cal L} &=&  - \frac{1}{{4{{e_{ph}} ^2}}}{F_{\mu \nu }}{F^{\mu \nu }} + \frac{1}{{12}}{H_{\mu \nu \lambda }}{H^{\mu \nu \lambda }} + \frac{{m{g_{ph}}}}{{2\pi }}{A_\mu }{F^\mu } \nonumber\\
&-& \frac{{6{m^2}g_{ph}^2}}{{\ln \frac{{{\Lambda ^2}}}{{{m^2}}}}}{F_\mu }\frac{1}{\Delta }{F^\mu } + \chi {\partial _\mu }{F^\mu }, \label{BF125}
\end{eqnarray}
where $\chi$ is the Lagrange multiplier to take into account the Bianchi identity. Whereas $\frac{1}{{e_{ph}^2}} = {e^2}\left( {1 + \frac{{{e^2}}}{{12{\pi ^2}}}\ln \frac{{{\Lambda ^2}}}{{{m^2}}}} \right)$ and ${g_{ph}} = \frac{g}{{6\pi }}\ln \frac{{{\Lambda ^2}}}{{{m^2}}}$. Here, we have made use of the same notation as in Ref. \cite{Diamantini}.

Before we proceed further, we shall pause to mention that the non-local ${F_\mu }\frac{1}{\Delta }{F^\mu }$ - term, which arises from the coupling between  the fermionic spin current and the rank - $2$ field, ${B_{\mu \nu }}$,
\begin{equation}
g{B_{\mu \nu }}{J^{\mu \nu }} =  - \frac{g}{2}{F^\mu }\frac{m}{\Delta }{J_\mu }, \label{BF126}
\end{equation}
may actually be seen as corresponding to a non-minimal coupling between the fermion and the dual of the field-strength ${H_{\mu \nu \kappa}}$. We could, from the very beginning, had started off with the non-minimal coupling accommodated in the $U(1)$- electromagnetic covariant derivative
\begin{equation}
{D_\mu }\Psi  \equiv {\partial _\mu }\Psi  + ie{A_\mu }\Psi  + i\frac{g}{m}{F_\mu }{\gamma _5}\Psi, \label{BF127} 
\end{equation}
with $g$ dimensionless. This however would not generate the ${F_\mu }{F^\mu }$-term in its non-local form      ${F_\mu }\frac{1}{\Delta }{F^\mu }$, which is crucial for the confining behavior of the potential, as we shall see more explicitly below. We then agree with the choice of the non-local formulation, as the authors of Ref. \cite{Diamantini} propose.

Next, in the same way as was done in the previous section, after integrating out the $B_{\mu\nu}$ field in favor of the $A_{\mu}$ field, the effective Lagrangian density reduces to
\begin{equation}
{\cal L} =  - \frac{1}{{4e_{ph}^2}}{F_{\mu \nu }}{F^{\mu \nu }} + \frac{1}{2}{\left( {\frac{{m{g_{ph}}}}{{2\pi }}} \right)^2}{A_\mu }\frac{\Delta }{{\left( {\Delta  + \gamma } \right)}}{A^\mu }, \label{BF130}
\end{equation}
where 
$\gamma  = \frac{{12{m^2}g_{ph}^2}}{{\ln \frac{{{\Lambda ^2}}}{{{m^2}}}}}$.

As in our preceding discussion, we now restore the gauge symmetry in equation (\ref{BF125}). This allows us to write the Lagrangian density as
\begin{equation}
L=  - \frac{1}{4}{F_{\mu \nu }}\left( {1 + \frac{{{\raise0.7ex\hbox{${{m^2}e_{ph}^2g_{ph}^2}$} \!\mathord{\left/
 {\vphantom {{{m^2}e_{ph}^2g_{ph}^2} {4{\pi ^2}}}}\right.\kern-\nulldelimiterspace}
\!\lower0.7ex\hbox{${4{\pi ^2}}$}}}}{{\left( {\Delta  + \gamma } \right)}}} \right){F^{\mu \nu }}. \label{BF135}
\end{equation}

Following the same steps that lead to (\ref{BF80}), the static potential for two opposite charges located at ${\bf y}$ and ${{\bf y}^ {\prime}}$ becomes
\begin{equation}
V =  - \frac{{{q^2}}}{{4\pi }}\frac{{{e^{ - ML}}}}{L} + {q^2}{\gamma}\ln \left( {1 + \frac{{{\Gamma ^2}}}{{{M^2}}}} \right)L, \label{BF140}
\end{equation}
where $L = |{\bf y} - {{\bf y}^ {\prime}}|$, ${M^2} ={m^2}g_{ph}^2\left( {{\raise0.7ex\hbox{$1$} \!\mathord{\left/
 {\vphantom {1 {4{\pi ^2}}}}\right.\kern-\nulldelimiterspace}
\!\lower0.7ex\hbox{${4{\pi ^2}}$}} + {\raise0.7ex\hbox{${12}$} \!\mathord{\left/
 {\vphantom {{12} {\ln {\raise0.7ex\hbox{${{\Lambda ^2}}$} \!\mathord{\left/
 {\vphantom {{{\Lambda ^2}} {{m^2}}}}\right.\kern-\nulldelimiterspace}
\!\lower0.7ex\hbox{${{m^2}}$}}}}}\right.\kern-\nulldelimiterspace}
\!\lower0.7ex\hbox{${\ln {\raise0.7ex\hbox{${{\Lambda ^2}}$} \!\mathord{\left/
 {\vphantom {{{\Lambda ^2}} {{m^2}}}}\right.\kern-\nulldelimiterspace}
\!\lower0.7ex\hbox{${{m^2}}$}}}$}}} \right)$ and $\Gamma$ is an ultraviolet cutoff. It is of interest also to notice that Lagrangians (\ref{BF125}) and (\ref{BF130}) are effective descriptions with cutoff $\Lambda$. So, our results are valid up to the energy scale $\Lambda$. Now, the potential (\ref{BF140}) must also be restricted to the same cutoff ($\Lambda$); therefore, it is sensible to identify the cutoff $\Gamma$, which appears in the derivation of the potential, with the cutoff $\Lambda$ ($ \Lambda$, $\Gamma$ $\gg$ $m$). Thus, we finally obtain that the static potential is given by
\begin{equation}
V =  - \frac{{{q^2}}}{{4\pi }}\frac{{{e^{ - ML}}}}{L} + {q^2}{\gamma}\ln \left( {1 + \frac{{{\Lambda ^2}}}{{{M^2}}}} \right)L. \label{BF141}
\end{equation}
The above static potential profile displays the conventional screening part, encoded in the Yukawa potential, and the linear confining potential. Accordingly, one of the most startling predictions of the interaction Lagrangian (\ref{BF120}) is the existence of a confining potential. Incidentally, it is of interest to notice that in the limit $m \to 0$ the confinement  disappears, which clearly shows the key role played by the fermions. It may be noted here that the confining potential of this model has been reported before \cite{Dutta}. However, in spite of their relevance, this result was obtained in a gauge-fixed scheme, and we think that this result should be corroborated by a gauge independent analysis.

Interestingly enough, the above static potential profile is analogous to that encountered in a theory of antisymmetric tensor fields that results from the condensation of topological defects as a consequence of the Julia-Toulouse mechanism \cite{GaeteWot1,GaeteWot2}. It is worth recalling here that the Julia-Toulouse mechanism is a condensation process dual to the Higgs mechanism proposed in \cite{Quevedo}, which describes phenomenologically the electromagnetic behavior of antisymmetric tensors in the presence of magnetic-branes (topological defects) that eventually condensate due to thermal and quantum fluctuations. Exploiting the previous phenomenology, we have studied in \cite {GaeteWot1,GaeteWot2} the dynamics of the extended charges (p-branes) inside the new vacuum provided by the condensate. More specifically, in \cite {GaeteWot1} we have considered the topological defects coupled both longitudinally and transversally to two different tensor potentials, $A_p$ and $B_q$, such that $p+q+2=D$, where $D=d+1$ space-time dimensions.
To be more precise, after the condensation the phenomenological Lagrangian density \cite{GaeteWot1} is given by
\begin{eqnarray}
{\cal L} &=& \frac{{{{\left( { - 1} \right)}^q}}}{{2\left( {q + 1} \right)!}}{\left[ {{H_{q + 1}}\left( {{B_q}} \right)} \right]^2} 
+ e{B_q}{\varepsilon ^{q,\alpha ,p + 1}}{\partial _\alpha }{\Lambda _{p + 1}} \nonumber\\
&+& \frac{{{{\left( { - 1} \right)}^{p + 1}}}}{{2\left( {p + 2} \right)!}}{\left[ {{F_{p + 2}}\left( {{\Lambda _{p + 1}}} \right)} \right]^2} 
\nonumber\\
&-& \frac{{{{\left( { - 1} \right)}^{p + 1}}\left( {p + 1} \right)!}}{2}{m^2}\Lambda _{p + 1}^2,  \label{BF145}
\end{eqnarray}
which reveals a $B$$\wedge$$F$ type of coupling between the $B_q$ potential with the tensor $\Lambda_{p+1}$ carrying the degrees of freedom of the condensate. In fact, following the procedure described in \cite{GaeteWot1}, we can further obtain the effective theory that results from integrating out the fields representing the vacuum condensate in the manner
\begin{equation}
{\cal L} = \frac{{{{\left( { - 1} \right)}^{q + 1}}}}{{2\left( {q + 1} \right)!}}{H_{q + 1}}\left( {{B_q}} \right)
\left( {1 + \frac{{{e^2}}}{{\Delta  + {m^2}}}} \right){H^{q + 1}}\left( {{B_q}} \right). \label{BF150}
\end{equation}  
From equation (\ref{BF150}) it now follows that for, $p=1$ and $q=1$, the effective theory can be brought to the form
\begin{equation}
{\cal L} = -\frac{1}{{4}}{F_{\mu \nu}}\left( A \right)\left( {1 + \frac{{{e^2}}}{{\Delta  + {m^2}}}} \right)
{F^{\mu \nu}}\left( A \right). \label{BF155}
\end{equation}

It is a simple matter to verify that equation (\ref{BF150}) reduces to equation (\ref{BF135}). With this then, we now see that the model studied here (\ref{BF135}) may be considered as some kind of effective theory that incorporates automatically the contribution of the condensate of topological defects to the vacuum of the model.

It follows from the above discussion a new connection among different effective theories, which are of interest from the point of view of providing unifications among diverse models as well as exploiting their equivalence in explicit calculations, as we have illustrated in this work.

\section{Final Remarks}

In summary, we have considered aspects of screening and confinement for a $B \wedge F$ model with topological mass from gauging spin. It was shown that the static potential profile is the sum of a Yukawa and a linear potential, leading to the confinement of static external charges. To do this, once again we have exploited a key aspect for understanding the physical contents of gauge theories, that is, the correct identification of field degrees of freedom with observable quantities.
We point out that our analysis reveals that similar results are obtained in a theory of antisymmetric tensor fields that results from the condensation of topological defects as a consequence of the Julia-Toulouse mechanism. Finally, it should be emphasized that an interesting feature of the present approach is to provide connections among different models. As we have already noticed, these connections show us a new sort of ''duality'' among diverse models and allow us to use this equivalence in concrete calculations, as we have illustrated in the present work.

\section{ACKNOWLEDGMENTS}
One of us (P. G.) wishes to thank the Field Theory Group of the COSMO/CBPF for the hospitality and the PCI-BEV/MCTIC support. P. G. was partially supported by Proyecto Basal FB 0821.

\end{document}